\documentclass[twocolumn,preprintnumbers,amsmath,amssymb]{revtex4}

\usepackage{graphicx}
\usepackage{dcolumn}
\usepackage{bm}
\usepackage{hyperref}
\usepackage{latexsym}
\usepackage{epsf}
\usepackage{amssymb}
\usepackage{amsmath}
\usepackage{slashed,epsfig}
\usepackage{bbm}
\usepackage{bbm}
\usepackage{amsmath,amssymb,amsthm}
\usepackage{pst-all}
\usepackage{xcolor}
\usepackage{booktabs}
\definecolor{Burgundy}{RGB}{155,5,8}
\hypersetup{
    colorlinks=true,
    linkcolor= Burgundy,
	citecolor = Burgundy,
    filecolor=magenta,      
    urlcolor=Burgundy,
    }
\newcommand{\be}{\begin{equation}}
\newcommand{\ee}{\end{equation}}

\begin{document}

\title{\textbf{Experimental tests of dark bubble cosmology}}

\author{U. Danielsson$^{1}$ and D. Panizo$^{1}$}
\affiliation{$^1$ Institutionen f\"{o}r Fysik och Astronomi, Box 803, SE-751 08 Uppsala, Sweden }

\begin{flushright}
	UUITP - 34/23
\end{flushright}

\begin{abstract}
\textbf{Abstract}: In this paper we use $\alpha'^2$-corrections to the $D_{3}$-brane action to obtain a non-zero and positive cosmological constant in the induced cosmology living on the dark bubble. Its measured value, together with the value of the fine structure constant, correspond to a dark dimension of size $5 \times 10^{-5}$ m and a string scale at $11$ TeV. We conclude that the dark bubble model predicts deviations of Newtonian gravity and stringy excitations of known particles within reach of future experiments.
\end{abstract}

\maketitle

\section{Introduction}\label{sec:Introduction}
\vspace{-0.2cm}

The construction of four-dimensional vacua with a tiny and positive cosmological constant (CC) $\Lambda$, is a big challenge for string phenomenology \cite{Danielsson:2018ztv, Obied:2018sgi, vanbeest2021lectures}. The dark bubble model proposal \cite{Banerjee:2018qey, Banerjee:2019fzz, Banerjee:2020wix, Banerjee:2020wov, Danielsson:2022fhd, 10DEmbedding, Basile:2023tvh} is an alternative framework that circumvents many of the problems that arise in standard compactifaction constructions. The key ingredient of the model is an unstable five dimensional anti-de Sitter space ($AdS_5$), which decays to a more stable one through the nucleation of a spherical $D_3$-brane. The induced metric corresponds to a four-dimensional expanding cosmology with a positive cosmological constant. For a recent review, where many common questions about the model are discussed, see \cite{FAQDark}.

In \cite{10DEmbedding},  the dark bubble was realized within type IIB string theory using a stack of $N$ $D_3$-branes, whose near horizon geometry takes the form of $AdS_{5} \times S^{5}$, with angular momentum in the compact directions. It has a holographic interpretation through $\mathcal{N} = 4$ super-Yang-Mills (SYM) theory with non-zero chemical potential $\mu$ and temperature $T$. From a five-dimensional point of view, the stack of branes can be viewed as an $AdS_5$-Reissner-Nordström (RN) black hole. Due to the instabilities caused by the presence of $\{\mu, T\}$, the black hole can discharge itself by emitting $D_3$-branes. The nucleation of a brane corresponds to the creation event of Vilenkin quantum cosmology as described in \cite{Danielsson:2021tyb}.

At zeroth order, the tension $\sigma$ of such branes is exactly critical and the cosmological constant vanishes, \footnote{According to the dark bubble proposal, dark energy is realised on the boundary of the bubble if the tension of the nucleated brane is \textit{subcritical}, i.e. $\rho_{\Lambda} = \sigma_{cr} - \sigma$ with $\sigma = \sigma_{cr} (1-\epsilon)$.}. In \cite{10DEmbedding}, it was argued that $\tfrac{1}{N}$ corrections to the tension of the brane will reduce this below the critical one, leading to an accelerated expansion with a positive cosmological constant. This in line with the Weak Gravity Conjecture (WGC)\cite{Arkani_Hamed_2007, ooguri2017nonsupersymmetric}. 

The goal of this paper is to calculate the precise size of such corrections, and relate the induced four-dimensional cosmological constant  to other fundamental scales. Remarkably, we will find that the theory is fully specified through the value of the cosmological constant together with the fine-structure constant. It uniquely predicts the string scale to be of order $10$ TeV, and the presence of large extra dimensions leading to deviations of the force of gravity on scales of order $10^{-5}$m.

\section{New energy hierarchy and $\mathbf{\Lambda_{4} >0}$ from string theory}\label{sec: new_hierarchy}

\subsection*{The energy hierarchy of the dark bubble}\label{sec: Nrule}

 We start by relating the various scales of the bulk theory. Our definitions are such that the relation between Newton's constant, reduced Planck mass and length in a $d$-dimensional space is:
\begin{equation}\label{eq: Planck mass}
    \kappa_{d} = 8 \pi G_{d} = M_{d}^{2-d} = l_d^{d-2}.
\end{equation} 
Throughout the paper we are working in units such that $\hbar =c=1$. In 10D super-gravity, the relation between the Planck length and the Regge slope parameter $\alpha'$ is : 
\begin{equation}\label{eq: l10_and_ls}
    l_{10}^8  = 2^{6} \pi^{7} g_s^2 \alpha'^{4}.
\end{equation}
Compactifying on $S^{5}$, we find a 5D Planck length given by
\begin{equation}\label{eq: comp_10_5}
    l_5^3 = \frac{l_{10}^{8}}{\text{Vol}(S^{5})} = 2^{6} \pi^{4} g_{s}^{2} \frac{\alpha'^{4}}{L^{5}} \,.
\end{equation}
In addition, the AdS/CFT correspondence \cite{Maldacena:1997re} states that the length scale $L$ of a stack of $N$ classical $D_3$ branes in type IIB supergravity is: 
\begin{equation}\label{eq: AdS_correspondence}
    L^4 = 4 \pi g_s N \alpha'^{2} \,,
\end{equation}
Using this, we can rewrite eq (\ref{eq: comp_10_5}) as:
\begin{equation}\label{eq: G5_and_L}
    L^3 = \frac{N^{2} l_{5}^{3}}{4 \pi^{2}} \, .
\end{equation}
The physical relevant regime has $g_{s} N \gg 1$, implying the following hierarchy of scales:
\begin{equation}\label{eq: first_hierarchy}
    \underbrace{\left(N/2 \pi \right)^{2/3} l_5}_{L} \gg\underbrace{\pi^{3/8} \left(N/2\pi\right)^{5/12}}_{l_{10}}\gg l_{5}.
\end{equation}
Next, let us  discuss how the junction conditions relate the induced four-dimensional cosmology and the bulk five-dimensional geometry \cite{Israel:1966rt}. The expanding bubble of vacuum is localised in the radial direction of the $AdS$ throat, dividing the bulk space in two: the true (inside) vacuum and the false (outside) vacuum. This generates a discontinuity of the bulk metric across the $D_3$ brane, which forces the presence of a non-zero induced energy-stress tensor on the brane $S_{ab}$, given by the second Israel's junction condition as:
\begin{equation}\label{eq: second_Israel}
    S_{ab} = \kappa_{5}^{-1} \left(\Delta K_{ab}- \Delta K h_{ab}\right)\big|_{+}^{-},
\end{equation}
where $K_{ab}$ is the extrinsic curvature of the brane defined as $K_{ab} = \nabla_{\beta}n_{\alpha}e^{\alpha}_{a}e^{\beta}_{b}$. The unit vector normal to the brane, $n_{\alpha}$, is defined in the direction of increasing transverse volume. The tangent vectors are defined by $e^{\alpha}_{a} = \partial x^{\alpha}/\partial y^{a}$, with $x^{\alpha}$ labeling bulk coordinates and $y^a$ labeling coordinates on the brane. In order for a spherical bubble to be able to nucleate, the tension of the bubble (i.e. $S_{ab} = \sigma\, h_{ab}$) can be at most the critical value given by
\begin{equation}\label{eq: tension_cr}
\sigma_{cr}=\frac{3}{8\pi G_5} \Delta k,
\end{equation}
where $\Delta k = \left(k_- - k_+\right)$, with $k_{\pm} = L^{-1}_{\pm}$. If the tension is less than the critical value, the difference will give rise to a positive cosmological constant \cite{Banerjee:2018qey}. The effective 4D gravitational constant can also be derived using the junction conditions and is given by 
\begin{equation}\label{eq: Newtons_DB}
    G_4 = \frac{2k_{-}k_{+}}{\Delta k} G_{5}.
\end{equation}
Here, one should note the crucial difference between the Randall-Sundrum construction, where the bubble has two insides, and the dark bubble where there is an inside and outside. With two insides, the extrinsic curvatures have the same sign, and the denominator of (\ref{eq: Newtons_DB}) is replaced by $k_-+k_+$. In case of RS, or any standard compactification, the higher dimensional gravity will be {\it stronger} than the 4D one. The dark bubble is unique since $\Delta k$ small allows the higher dimensional gravity to be {\it weaker}. We will come back to this later.   

The relations between bubble boundary and bulk, (\ref{eq: tension_cr}, \ref{eq: Newtons_DB}), that we have obtained, can be expressed in terms of parameters of the 10D space time. We first notice that $G_{5}$ in expression (\ref{eq: G5_and_L}) per definition remains constant across the junction (i.e. $\Delta G_{5} = 0$), which implies:
\begin{equation}\label{eq: Nk-expression}
    \Delta k = -\frac{2}{3} \frac{\Delta N}{N} k.
\end{equation}
This expression relates the scales $k_{\pm}$ of the outside/inside vacua to the $\Delta N$ number of $D_3$-branes that can potentially nucleate and escape from the stack. This allows us to write (\ref{eq: tension_cr}) as:
\begin{equation}
    \sigma_{cr}=-\frac{k}{4\pi G_5} \frac{\Delta N}{N}= - \frac{\Delta N}{(2\pi)^3 g_s \alpha '^2} = - \Delta N T_{D_3},
\end{equation}
where $T_{D_3} = ((2\pi)^3 g_s \alpha '^2)^{-1} $ is the tension of a fundamental $D_3$-brane and we have again used eq(\ref{eq: AdS_correspondence}). If we assume the nucleation of a single brane, then $\Delta N=-1$, so we find that
\begin{equation}\label{eq: crit_T3}
    \sigma_{cr} = T_{D_3}.
\end{equation}
This is {\it exactly} the correct tension for a single BPS $D_3$-brane, and we conclude that such a nucleated brane has critical tension $\sigma_{cr}$. Below we will see how corrections reducing the tension of the brane will generate a positive cosmological constant (i.e. $\rho_{\Lambda} = \sigma_{cr} - \sigma >0$) on the induced four-dimensional cosmology.

We will now write the value of the 4D Newton's constant in terms of the $N$ number of branes in the background. Making use of expression (\ref{eq: Nk-expression}), one can rewrite (\ref{eq: Newtons_DB}) as:
\begin{equation} \label{eq: G4G5}
G_4=2\frac{k_- k_+}{k_- - k_+} G_5 =\frac{3 k^2}{-\Delta N k}N G_5 =3\frac{N}{L} G_5 \, , 
\end{equation}
where we have written $k_{-} \simeq k_{+}$, with $k_{\pm} = k \mp \Delta k$ and $\Delta N = -1$. Using the previous relations, we can then connect $L$ to the 4D Planck scale through
\begin{equation}\label{eq: Ldpendingl4}
    L = \frac{N^{1/2}}{2\sqrt{3} \pi} l_4,
\end{equation}
which allow us to rewrite the hierarchy (\ref{eq: first_hierarchy}) in terms of $l_4$ instead:
\begin{equation}\label{eq: second_hierarchy}
    \underbrace{\frac{N^{1/2}}{2 \sqrt{3}\, \pi} \,l_4}_{L} \:\gg\: \underbrace{\frac{N^{1/4}}{2^{3/4} \sqrt{3} \,\pi^{3/8}} \, l_{4}} _{l_{10}} \:\gg\: l_4 \:\gg\: \underbrace{\frac{N^{-1/6}}{\sqrt{3} (2\pi)^{1/3}} \, l_{4}}_{l_{5}}. 
\end{equation}
As already pointed out, it is of crucial importance to notice how unusual the hierarchy $l_{5} \ll l_{4}$ is. In conventional dimensional reduction, $L \gg l_{5}$ immediately leads to $l_{5} \gg l_{4}$, which implies that gravity is stronger in the compact dimensions. As a consequence, there is an upper bound on the volume of those closed directions to avoid conflict with observations. This does no longer apply to the dark bubble model. The reason is the presence of the large factor $N$ in the relation between the 5D and 4D Planck scale in the expression (\ref{eq: second_hierarchy}), provided by the derivation of four-dimensional Newton's constant from the junction conditions. 

Let us now turn to the expected corrections to the BPS tension of the $D_3$-brane mediating the decay.

\subsection*{$\mathbf{\Lambda_{4} > 0}$ from subleading corrections}

As discussed above, eq(\ref{eq: crit_T3}) implies the absence of an induced cosmological constant at leading order in the brane action. As argued in \cite{10DEmbedding}, it is the angular momentum of the stack of branes (or the high chemical potential and non-zero temperature in dual field theory) that breaks the supersymmetry of the system. This breaking should be reflected by corrections to the brane's tension, making it sub-critical, in line with the WGC \cite{Arkani_Hamed_2007} and the decay of non-supersymmetric AdS vacua described in \cite{ooguri2017nonsupersymmetric}. These corrections are expected to be of the order $1/N$, as the nucleation of a $D_3$-brane can be understood as a Higgsing process of the dual gauge group as $SU(N) \rightarrow SU(N-1) \times U(1)$, \cite{10DEmbedding}.

There are two types of corrections (up to a power of the string coupling) that can roughly contribute at the same order of magnitude $\tfrac{1}{N}$. Let us elaborate on them.

\begin{itemize}
    \item \textbf{"Stringy" corrections}: These corrections at $\mathcal{O}(\alpha'^{2})$ can be independently computed for the Dirac-Born-Infeld (DBI) and Wess-Zumino (WZ) pieces of the brane action. Curvature corrections to the DBI action were found in \cite{Bachas:1999um} by requiring consistency of the effective action with the $\mathcal{O}(\alpha'^{2})$ terms of the corresponding disk-level scattering amplitude \cite{Garousi_1996}. On the other hand, the WZ curvature corrections can be found by imposing that the chiral anomaly on the world volume of intersecting D-branes cancels with the anomalous variation of the WZ action \cite{Green_1997}. We will explore each of them in the following section. Schematically, we can write these corrections as:
        \begin{equation}
            \delta \sigma \sim T_{D_3}\,\frac{\alpha'^{2}}{L^4}\sim T_{D_3}\,\frac{1}{g_s N},
        \end{equation}
    where we have used relation (\ref{eq: AdS_correspondence}). Observe that, since the length scale $l_{s}^{2} = \alpha'$ is a classical scale set by the tension of the string, these contributions can be interpreted as a tree level effect.
    
    \item \textbf{Quantum loop corrections}: Quantum loop corrections can be written as:
        \begin{equation}
            \frac{1}{N} \sim \frac{l_{10}^4}{L^4} ,
        \end{equation}
    where $l_{10}$ is the 10D Planck length. The Planck length is obtained from the classical Newton's constant by introducing Planck's constant, showing that this correction comes from quantum loops. A renormalized one-loop calculation for a massive field of mass $m$, leads to a contribution to the vacuum energy of order $m^4$. Hence, we see that the loops above would be generated by fields with $m\sim 1/L$, which are present in the theory. \footnote{Note that this also is the mass-scale of the neutrino.}
\end{itemize}
While stringy corrections to the brane action will be suppressed by a factor $1/g_s N$, quantum loop corrections will be suppressed by $1/N$. This implies that stringy corrections will be the most important, provided that $g_{s} \ll 1$. We conclude that corrections from string theory will induce a 4D cosmological constant according to:
\begin{equation}
    \rho _\Lambda \sim T_{D_3}\, \frac{1}{g_s N} \sim \frac{1}{g_s L\,G_5\,N^2} \sim \frac{1}{g_s L^4} .
\end{equation}
In the next section, we will compute in detail all relevant contributions to $\Lambda_{4}$ from higher curvature corrections to the brane action.

\section{Stringy corrections to the brane action} \label{sec: corrections}

The low energy effective field theory of $D_{p}$-branes in type II superstring theories consists of the DBI \cite{Bachas_1996DBI} and the WZ \cite{douglas1995branesWZ} actions. Concretely, for $p=3$, this is:
\begin{equation}\label{eq:D_3action}
  S_{D_3} = -T_{D_3}\int d^4\xi \sqrt{-\det P[G]}+T_{D_3}\int P[C_4],
\end{equation}
where $P[\,\cdot\,]$ denotes the pullback of a 10D space-time field to the brane world-volume. $G$ is the 10D metric and $C_{4}$ is the four-form that charges the brane. From now on, we will refer to ten dimensional directions with Greek indices, $\{a, b, c, ...\}$ to denote space-time coordinates on the brane and $\{i, j, k,...\}$ for directions normal to this.

At leading order with critical tension, the two terms in (\ref{eq:D_3action}) are perfectly balanced, so that they cancel. We have previously commented that we expect corrections in line with WGC making the tension slightly less than the critical value. It is natural for such corrections to appear in the DBI-part, but not in the WZ-part. The latter represents the change in the bulk energy density between the inside and the outside, and can be viewed as a boundary term appearing in a bulk calculation. We do not expect this to change through local effects having to do with the brane. Hence, the WZ-part should not give any correction relevant for us. We will find evidence supporting this claim below. 

\subsection*{Corrections to DBI}

String theory corrections to the DBI-action at order $\alpha '^2$ due to curvature were obtained in \cite{Bachas:1999um}:
\begin{equation}
\begin{aligned}
     &S_{\rm DBI}\supset -T_{D_{3}} \frac{\pi^2 \alpha '^2}{48} \int d^4 x\,  e^{-\Phi} \sqrt{-g} \big ( \, (R_T)_{abcd} (R_T)^{abcd}-\\
     &-2(R_T)_{ab} (R_T)^{ab}-(R_N)_{abij} (R_N)^{abij}+2(\bar{R})_{ij} (\bar{R})^{ij}\, \big).
\end{aligned}
\end{equation}
Observe that $(R_T)_{abcd}=R^{(4)}_{abcd}$ is the induced curvature on the brane and is related to the bulk curvature $R^{(4+d)}_{\mu \nu \rho \sigma}$ and the extrinsic curvature $K^i _{ab}$, with $i=\left\{1 ... d \right\}$ through the Gauss-Codazzi (GC) equations:
\begin{equation}
R^{(4+d)}_{\mu \nu \rho \sigma} e^\mu_a e^\nu_b e^\rho_c e^\sigma_d = (R_T)_{abcd} + \delta_{ij}\left(K^i _{ad} K^j _{bc} - K^i _{ac} K^j _{bd}\right),
\end{equation}
where $e^\mu_a$ projects the Riemann tensor of the bulk onto the brane. The number of dimensions normal to the brane is represented by $d$. In our case, the bulk will be $AdS_5 \times S^5$, so $d=6$, \footnote{As we will see later the $S^5$ will not contribute at leading order and we will only have the direction $z$ transverse to the brane in AdS$_5$ to worry about. The extrinsic curvature then reduces to just $K_{ab}$.} $(R_N)_{ab}^{\quad ij}$ is the curvature of the normal bundle, and given through
\begin{equation}
R^{(4+d)\: ij}_{\mu \nu} e^\mu_a e^\nu_b =(R_N)_{ab}^{\quad ij}  + h^{cd} \left( K^i _{ad} K^j _{bc} - K^i _{ac} K^j _{bd}\right).
\end{equation}
Finally, we need to define
\begin{equation}\label{eq: Rij}
 \bar{R}^{ij}= h^{ab} R^{(4+d) \: ij}_{ab}+ h^{ab} h^{cd} \left(K^i _{ac} K^j _{bd} - \eta K^i _{ab} K^j _{cd}\right).
\end{equation}
In contrast to the other corrections to the action, expression (\ref{eq: Rij}) has no clean geometrical interpretation. We note two terms depending on the square of the extrinsic curvature. The first one is of the form $K^i _{ac} K^{ac \:j}$, and was derived in \cite{Bachas:1999um}. The second term, of the form $K^{a\: i} _{a} K^{b \:j}_b$, has an unknown coefficient $\eta$, which could not be determined through the scattering amplitudes in \cite{Bachas:1999um}.  These amplitudes involve gravitons and scalar perturbations of flat space that are on-shell. Since $K _{ab}^i= \partial_a \partial _b \phi^i$ in flat space, with $\phi ^i$ as the embedding coordinate, it follows that the trace of $K$ must vanish. Therefore, a calculation using scattering amplitudes in flat space does not fix the constant $\eta$ above.

Unfortunately, it is exactly this term that will yield a non-trivial correction to the DBI-action of our dark bubble, embedded into AdS-space. A way to fix $\eta$, as well as many other parameters when more fields are turned on, is to use T- and S-dualities. This was done in \cite{Garousi:2015} with the result that $\eta=1$, if S-duality is used off-shell. However, the derivation is slightly subtle. The sum of the contact terms in the scattering amplitudes, which are used to read off the terms in the effective action, should respect T-duality off shell.  S-duality, on the other hand, may mix the contact terms with the pole terms, which makes it much more tricky to use. Actually, since on-shell S-duality is automatically satisfied by the scattering amplitudes, it is really only off-shell S-duality that needs extra care. Luckily, for precisely those amplitudes that are relevant for this paper, no such mixing occurs. This is the reason why we can trust the derivation in \cite{Garousi:2015}, based on off-shell T-duality, and use $\eta=1$, \footnote{We thank M. Garousi for discussions and explanations.}. Clearly, it would be satisfying to derive this result through an explicit scattering amplitude calculation. This could be achieved considering three closed string gravitons, scattering off a D-brane. You would then induce a vertex between one of the gravitons and two internal off-shell scalars. Such calculations turn out to be extremely complicated, but can in principle be done, \footnote{We thank Oliver Schlotter and Stephan Stieberger for discussions.}.
 
We will now use the correction described above to compute the shift in the tension of the brane, and thus the cosmological constant of the dark bubble. To do this, it is enough for us to study a pure $AdS_5$-background. We are only interested in finding the value of the cosmological constant, and therefore we consider the late time universe at large radius, where matter components induced by the Reissner-Nordström piece of the metric of \cite{10DEmbedding, Henriksson:2019zph} can be ignored. For a cosmological constant that is small compared to all other scales of the model, we can ignore the expansion of the dark bubble when calculating its value. We therefore assume that the brane sits at a constant $z=z_0$, and we can also assume that the brane sits at a constant position on the $S^5$. With $AdS_5 \times S^5$ as a direct product, and the brane positioned within the AdS$_5$, all contributions to the extrinsic curvature $K_{ab}^i$ with $i\neq z$ will vanish. With a single extra dimension to worry about we just get $(R_N)_{ab}^{\quad zz}=0$ by antisymmetry. Since the brane is flat, $R_T$ will also vanish. One may also note that the expression for 5D Riemann tensor of AdS, given by
\begin{equation}
    R^{(5)}_{abcd}= k^2 \left( h_{ad} h_{bc}- h_{ac} h_{bd} \right),
\end{equation}
with $h_{ab}=g_{ab}$,  coincides with the Gauss-Codazzi equations for a flat brane, with extrinsic curvature
\begin{equation}
    K_{ab} = k h_{ab} .
\end{equation}
We can now evaluate the single non-zero component of $ \bar{R}^{ij}$ to obtain
\begin{equation}
    \bar{R}^{zz}=k^2 g^{zz} \left( -4 +4 -16\right) = -16 k^2 g^{zz} .
\end{equation}
If we then sum up all contributions, we find
\begin{equation}
    \begin{split}
        &(R_T)_{abcd} (R_T)^{abcd}-2(R_T)_{ab} (R_T)^{ab}-\\
        -&(R_N)_{abij} (R_N)^{abij}+2(\bar{R})_{ij} (\bar{R})^{ij} = 512 k^4 .
    \end{split}
\end{equation}
This will make the tension {\it smaller}, just as expected from WGC. This means:
\begin{equation}\label{eq: tension_shift}
    T_{D_{3}} \rightarrow T_{D_{3}} \left(1-\frac{512}{48} \pi^2 \alpha '^2 \, k^4 \right).
\end{equation}

\subsection*{Corrections to WZ}

We have already argued that the WZ-part of the action will not yield any non-vanishing correction in our case. Let us look at this in more detail. It describes the coupling of the brane to Ramond-Ramond (RR) fields of various dimensions. It is given by:
\begin{equation}
    S_{WZ}= T_{D_p} \int d^{p+1} x \, C \wedge\sqrt{\frac{{\cal A}(4\pi^2 \alpha  '^{2} R_T)}{{\cal A}(4\pi^2 \alpha'^{2} R_N)}}.
\end{equation}
Since $R_T$ and $R_N$ all vanishes for us, there are no relevant corrections form this expression. However, as shown in \cite{Garousi:2016}, there is also a related contribution to the WZ-action given by
\begin{equation} \label{eq: WZ corr}
    \begin{split}
        S_{WZ} \supset -\frac{\pi^2 \alpha'^{2}}{12} \int &d^4 x e^{-\Phi} \sqrt{-g}\epsilon^{a_0 a_1 a_2 a_3}\\ 
        &\tfrac{1}{(p+1)!}\partial _z {\cal F}^{(5)}_{z a_0 a_1 a_2 a_3} \bar{R}^{zz}.
    \end{split}
\end{equation}
Here we again note the appearance of $\bar{R}^{zz}$, which contains the trace of the extrinsic curvature. In contrast to the corresponding result for the DBI-action, the results in \cite{Garousi:2016} are not conclusive. A reasonable guess is that $\bar{R}^{zz}$ is the same as for the DBI part, but the dualities cannot fully fix it. Luckily, as we have already explained, there are good reasons to expect that we will not need the precise result, and that the contribution vanishes for a very general reason. Studying (\ref{eq: WZ corr}), we note the presence of $\partial _z {\cal F}^{(5)}_{z a_0 a_1 a_2 a_3}$. In the analysis of \cite{Garousi:2016}, assuming the trace of the extrinsic curvature to vanish, there is no difference between $\partial$ and the full covariant derivative $\nabla =\partial +...$. One could therefore argue that the correct expression should be (\ref{eq: WZ corr}) with a {\it covariant} derivative. Then the correction to the WZ will simply vanish as a result of the equations of motion for ${\cal F}$. This is in line with our general expectation.

\section{Phenomenological implications and conclusion}\label{sec: conclusions}

The results we have obtained in this paper are not just order of magnitude estimates, but based on first principles in the precise model put forward in \cite{10DEmbedding}. Let us now fix the scales and parameters by comparing with cosmological observations.

From expression (\ref{eq: tension_shift}) and $\rho_{\Lambda} = \sigma_{\rm cr} - \sigma$, we see that:
\begin{equation}
    \rho_{\Lambda} = \frac{32}{3} \frac{\pi^{2} \alpha'^{2}}{L^{4}} \, T_{D_3}  \quad \rightarrow \quad \rho_{\Lambda} = \frac{4}{3\pi g_{s}\, L^{4}}  ,
\end{equation}
where eq(\ref{eq: AdS_correspondence}) and $T_{D_3} = ((2\pi)^3 g_s \alpha '^2)^{-1} $ have been used. Using equations (\ref{eq: Planck mass}), (\ref{eq: G5_and_L}) and (\ref{eq: G4G5}) we find
\begin{equation}
    L^2=\frac{2}{3\pi} N G_4  ,
\end{equation}
implying
\begin{equation}\label{eq: number_N}
    N = \frac{\sqrt{3 \pi}}{ G_4 \sqrt{g_{s} \rho_\Lambda}}.
\end{equation}

To constrain the value of the string coupling $g_{s}$, we notice that:
\begin{equation}
    M_{p} = M_{{\rm RN}},
\end{equation}
where 
\begin{equation}
    M_{p} = T L=\frac{1}{L} \sqrt{\frac{g_sN}{\pi}}=\sqrt{\frac{3 g_s}{2 G_4}},
\end{equation}
is the mass of a four-dimensional particle represented by the end point of a fundamental string with the fundamental charge $e$ and tension $T = 1/2\pi \alpha'$, attached to the $D_{3}$-brane \cite{Banerjee:2019fzz}. This BPS-particle will induce a Reissner-Nordström geometry. In the extremal case scenario, this results in $M_{{\rm RN}}$ to be:
\begin{equation}
    M_{{\rm RN}} = \sqrt{\frac{\alpha_{\rm EM}}{G_4}},
\end{equation}
where $\alpha_{\rm EM} \simeq \tfrac{1}{137}$, is the fine structure constant. This implies that the string coupling constant can be expressed in terms of the fine structure constant as:
\begin{equation}
    g_{s} = \frac{2}{3} \alpha_{\rm EM}.
\end{equation}
Electromagnetism is the only long range force, except for gravity, and it is therefore natural that the low energy value of the string coupling is set by it.
Introducing this result in expression (\ref{eq: number_N}), together with $\rho_\Lambda \simeq 6.8 \times 10^{-27} {\rm kg/m^3}$, we obtain (after re-inserting $\hbar$ and $c$):
\begin{equation}
    N = 1.2 \times 10^{63},
\end{equation}
which fixes the hierarchy of scales of the dark bubble construction as described in table (\ref{Table: 1}). One should note that $L$, $l_s =\sqrt{\alpha'}$, $G_5$, and $G_{10}$, as well as $g_s$ and $N$ are all unambiguously fixed. Throughout the calculations, we have used the reduced Planck scale for convenience. In the table we have chosen to instead use $G_d=\tilde{l}_d^{d-2}$.
\begin{table}[h!]
	\centering
	\begin{tabular}{c|c|c}
	\textbf{Scale} & \textbf{Length (m)} & \textbf{Energy} \\
	\hline
    \hline
	$L$ &  $5.1 \times 10^{-5}$ &  $3.8 \, {\rm meV}$ \\
    $\sqrt{\alpha'}$ & $1.8 \times 10^{-20}$ & $11.2\, {\rm TeV}$  \\
    $\tilde{l}_{10}$ & $1.4 \times 10^{-20}$ &  $13.7 \, {\rm TeV}$ \\
    $\tilde{l}_{5}$ & $3.9 \times 10^{-45}$ & $5.1 \times 10^{28} \,{\rm TeV}$
	\end{tabular}
	\caption{New hierarchy of scales associated to the dark bubble embedding into string theory.}
	\label{Table: 1}
\end{table}

The dark bubble model was introduced as a way to naturally incorporate a positive cosmological constant. It is remarkable that a detailed implementation of the model leads to precise predictions of novel phenomena in microscopic physics. The presence of the large extra dimensions implies a modification of gravity at scales of order a few $10^{-5}$m. Similar proposals have been made previously in the literature \cite{Montero:2022prj}, but our result is through calculations based on first principles. Furthermore, the way the extra dimensions relate to the 4D world is completely new. Moreover, the model predicts new high energy physics at energy scales of tens of TeV in the form of stringy excitations of known particles such as the photon.

\section*{Acknowledgments}
We thank Oscar Henriksson for discussions at an early stage of the project. We are also grateful to Mohammad R. Garousi for enlightening correspondence, and to Ivano Basile, Suvendu Giri, Oliver Schlotterer and Thomas van Riet for fruitful discussions. DP would like to thank the Centre for Interdisciplinary Mathematics (CIM) for financial support. 

\newpage

\renewcommand{\tt}{\normalfont\ttfamily}
\bibliography{main}
\bibliographystyle{utphys.bst}
\end{document}